\documentclass[prl,showpacs,twocolumn,preprintnumbers,amsmath,amssymb,superscriptaddress]{revtex4}

\usepackage{graphicx}
\usepackage{color}
\usepackage{dcolumn}
\usepackage{bm}
\usepackage{epsfig,float,afterpage,amssymb,wrapfig,psfrag}

\begin{document}
\title{Kondo destruction and valence fluctuations in an Anderson model}
\author{J.~H.~Pixley}
\affiliation{Department of Physics \& Astronomy, Rice University,
Houston, Texas, 77005, USA}
\author{Stefan Kirchner}
\affiliation{Max Planck Institute for the Physics of Complex Systems,
01187 Dresden, Germany}
\affiliation{Max Planck Institute for Chemical Physics of Solids,
01187 Dresden, Germany}
\author{Kevin Ingersent}
\affiliation{Department of Physics, University of Florida,
Gainesville, Florida 32611-8440, USA}
\author{Qimiao Si}
\affiliation{Department of Physics \& Astronomy, Rice University,
Houston, Texas, 77005, USA}

\date{\today}

\begin{abstract}
Unconventional quantum criticality in heavy-fermion systems has been extensively
analyzed in terms of a critical destruction of the Kondo effect. Motivated by a
recent demonstration of quantum criticality in  a mixed-valent heavy-fermion
system, $\beta$-YbAlB$_4$, we study a particle-hole-asymmetric
Anderson impurity model with a pseudogapped density of states. We demonstrate
Kondo destruction at a mixed-valent quantum critical point, where a collapsing
Kondo energy scale is accompanied by a singular charge-fluctuation spectrum.
Both spin and charge responses scale with energy over temperature ($\omega/T$)
and magnetic field over temperature ($H/T$). Implications for unconventional
quantum criticality in mixed-valence heavy fermions are discussed.
\end{abstract}


\maketitle

Competing interactions in quantum systems give rise to zero-temperature phase
transitions. If it is continuous, such a transition takes place at a quantum
critical point (QCP). There is mounting evidence, especially in heavy-fermion
systems, that a QCP can underlie unconventional superconductivity
\cite{Mathur.98}; related considerations have been applied to high-temperature
cuprate and iron pnictide superconductors \cite{Broun.08}.
It is standard to describe a QCP within the Ginzburg-Landau-Wilson (GLW)
framework: critical destruction of an order parameter characterizing a
spontaneously broken symmetry gives rise to collective modes associated with
order-parameter fluctuations \cite{Sachdev}. In the context of
antiferromagnetic metals, this is referred to as a spin-density-wave QCP
\cite{Hertz.76+Millis.93}.

Recent experiments in heavy-fermion metals have clearly established the
existence of a novel class of antiferromagnetic QCPs, characterized by non-Fermi
liquid behavior and $\omega/T$ scaling in the dynamical spin susceptibility
\cite{Loehneysen.07+Si.10+Aronson.95+Schroeder.00+Paschen.04+Friedemann.10}.
There are indications that such unconventional QCPs also promote
superconductivity \cite{Park.06}. These QCPs defy description in terms of a
GLW functional \cite{Si.01+Si.03,Coleman.01}; their understanding requires the
introduction of quantum modes beyond order-parameter fluctuations. The
proposed additional modes are associated with the critical destruction of the
Kondo effect \cite{Si.01+Si.03,Coleman.01}. In the paramagnetic phase, Kondo
singlets form and generate Kondo resonances, thereby turning the local
moments into single-electronic excitations and enlarging the Fermi
surface. The destruction of the Kondo effect across the antiferromagnetic QCP
suppresses the Kondo resonances, making the Fermi surface small. 
Critical Kondo destruction therefore manifests itself in a
discontinuous evolution of the Fermi surface across the transition, as has
been observed through quantum oscillation and Hall effect measurements
\cite{Loehneysen.07+Si.10+Aronson.95+Schroeder.00+Paschen.04+Friedemann.10,Shishido.05}.

Theoretical studies of critical Kondo destruction have largely been confined to
the Kondo-lattice limit of integer valence. In rare-earth intermetallics, 
superconductivity is believed also to arise in the vicinity of valence
transitions \cite{Yuan.03+Holmes.07+Rueff.11}, which have been found to be
first order. Until recently, there has been no significant evidence for
a QCP associated with valence fluctuations. The situation has changed with the
observation of mixed valency in the ytterbium-based heavy-fermion
superconductor $\beta$-YbAlB$_4$ \cite{Okawa.10}, which is quantum critical
under ambient conditions \cite{Nakatsuji.08}. In an applied magnetic field,
the magnetization obeys $H/T$ scaling \cite{Matsumoto.11}, consistent with the
$\omega/T$ scaling seen previously near the unconventional QCPs of
antiferromagnetic heavy-fermion compounds. These properties implicate
$\beta$-YbAlB$_4$ as a strong candidate for a mixed-valent heavy-fermion QCP,
and raise the prospect that the material's unusual scaling behavior can be
understood in terms of critical Kondo physics.

At first glance, critical Kondo destruction at mixed valence appears
unlikely. Kondo destruction in a Kondo lattice amounts to the
localization of $f$ electrons. While unconventional, this is physically
transparent, because localization can readily arise for a commensurate
filling of an electronic orbital (one $f$-electron per site). At mixed
valence, the situation is more subtle because the $f$ orbital has a
fractional, generally incommensurate, per-site occupancy, and there is no
mechanism known for electron localization at incommensurate fillings. 
This leads to important questions of principle: Can critical Kondo destruction
occur in the presence of valence fluctuations and, if so, how does the
criticality compare to its local-moment counterpart? For instance, are charge
excitations part of the critical fluctuation spectrum?

In this Letter, we address these issues in the mixed-valence regime of an
Anderson impurity model whose conduction-electron density of states features a
pseudogap centered on the Fermi energy. We focus on an impurity model because
of the local nature of the Kondo-destruction physics; formally, the Kondo
lattice model can be treated through an effective impurity model in the
extended dynamical mean field approach \cite{SmithSi-edmft,Chitra}.
Given that the commensurate-filling (i.e., local-moment) limit
of the model exhibits critical Kondo destruction and associated dynamical
scaling properties \cite{Ingersent.02}, we consider the pseudogapped density
of states to provide a prototype setting to search for a Kondo-destruction QCP
at mixed valence. Our model has the advantage of being amenable to study using
reliable methods: the continuous-time quantum Monte Carlo (CT-QMC)
method \cite{Gull.11} and the numerical renormalization group (NRG)
\cite{Bulla.97,Buxton.98,Ingersent.02}. 

Surprisingly, we do find critical Kondo destruction in this mixed-valent
model. The critical properties in the spin sector reflect the collapse of an
energy scale as the QCP is approached from the Kondo-screened side but not from
the Kondo-destroyed side, much as in the integer-valent (local-moment) limit.
By contrast, the charge sector shows a collapsing energy scale on both sides of
the QCP. The critical point displays $H/T$ (and $\omega/T$) scaling. This
existence proof for a Kondo destruction QCP at mixed valence makes it feasible
to interpret the $H/T$ and related scaling properties of $\beta$-YbAlB$_4$ in
terms of an interacting fixed point.
We note that the same model is also relevant to impurity physics in $d$-wave
superconductors and graphene, where the density of bulk fermionic states goes
to zero at the chemical potential \cite{Chen.11+Jacob.11}.

\begin{figure}[t!]
\includegraphics[width=\linewidth]{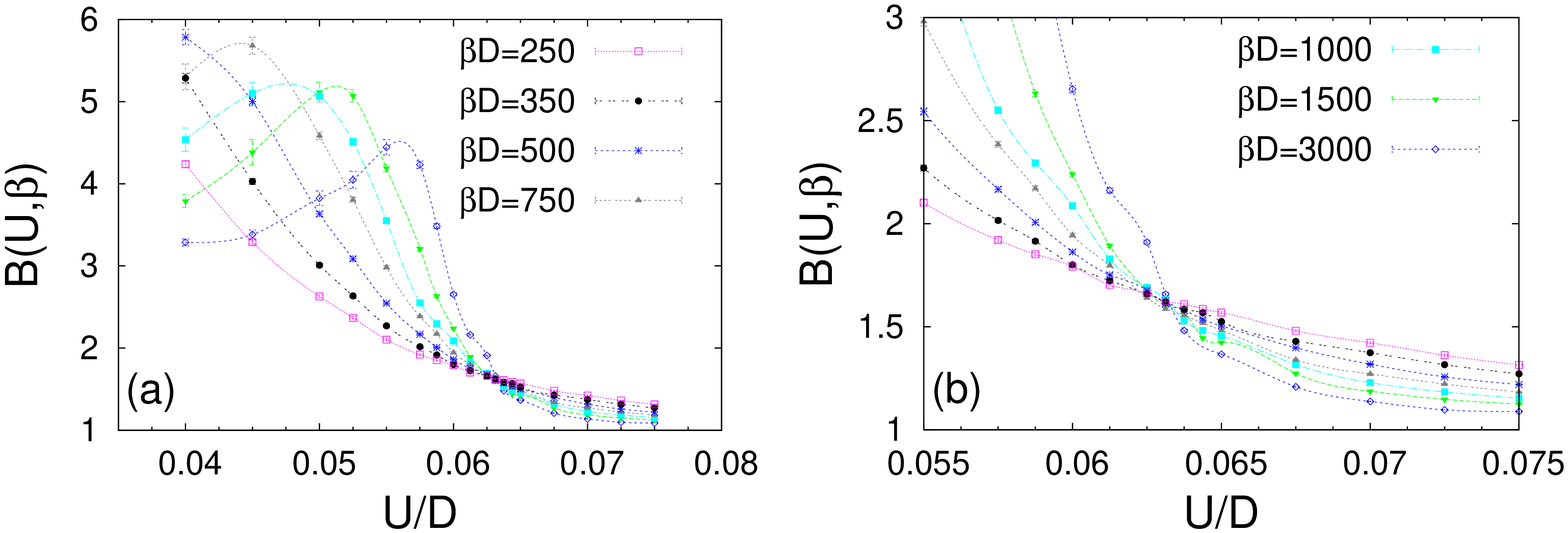}
\caption{(a) Binder cumulant $B(U,\beta)$ vs.\ $U$ for $r=0.6$, $\Gamma_0=0.1D$,
and $\epsilon_d=-0.05D$, and the labeled temperatures.
Error bars were obtained from a jackknife analysis of the CT-QMC data.
(b) Blow up of the same data around the intersection of curves, which
determines $U_c/D = 0.063125 \pm 0.0008$.}
\label{Figure1}
\end{figure}

The Anderson impurity Hamiltonian is
\begin{equation}
H = \sum_{k, \sigma} \bigl[ \epsilon_k c_{k\sigma}^{\dag} c_{k\sigma}
+ V \bigl(  d_{\sigma}^{\dag}c_{k\sigma} + \text{H.c.}\bigr) \bigr]
+ \varepsilon_d n_d  + U n_{d\uparrow}n_{d\downarrow} ,
\label{ham}
\end{equation}
where $c_{k\sigma}$ annihilates a conduction-band electron of energy
$\epsilon_k$, $d_{\sigma}$ annihilates an electron of energy $\varepsilon_d$
in the impurity level, $U$ is the electron-electron repulsion within the
impurity level, $V$ is the hybridization taken to be momentum independent,
$n_{d\sigma} = d^{\dag}_{\sigma}d_{\sigma}$, and
$n_d = n_{d\uparrow} + n_{d\downarrow}$.
The band density of states vanishes in a power-law fashion at
the Fermi energy ($\epsilon_F=0)$:
\begin{equation}
\rho(\epsilon) = \sum_{k}\delta(\epsilon - \epsilon_k)
= \rho_0|\epsilon/D|^r
\Theta(D - |\epsilon|).
\label{dos}
\end{equation}
The impurity-band interaction is completely specified by the imaginary part of
the hybridization function,
$\Gamma(\epsilon)=\pi\sum_k V^2 \delta(\epsilon-\epsilon_k)
=\Gamma_0|\epsilon/D|^r$,
where $\Gamma_0 = \pi \rho_0 V^2$.

The critical properties of the model with particle-hole (p-h) symmetry
($\varepsilon_d=-U/2$) and its Kondo limit ($U \gg \Gamma_0$, where
local charge fluctuations are negligible) have been investigated in a number
of analytic and numerical studies \cite{Buxton.98,Ingersent.02,Withoff.90+Vojta.02+Glossop.03+Fritz.06,Vojta.04,Glossop.11}.  
The breaking of p-h symmetry is irrelevant for pseudogap exponents $r$
in the range $0<r<r^{\ast} \simeq 0.375$, but becomes relevant for
$r> r^{\ast}$, leading to a mixed-valent QCP \cite{Buxton.98}; $r=1$ serves as
an upper critical ``dimension'', above which the critical properties have a
mean-field character \cite{Ingersent.02,Vojta.04}.  

Here, we investigate the p-h-asymmetric pseudogap Anderson model by varying $U$ 
for fixed $\Gamma_0$ and $\varepsilon_d$ to pass from a Kondo-screened
strong-coupling phase ($U<U_c$) to a Kondo-destroyed local-moment phase
($U>U_c$). We apply the CT-QMC technique, which was recently shown to be able
to reach temperatures $T$ sufficiently low to access the quantum critical
regime \cite{Glossop.11}. We measure the dynamical local spin and charge
susceptibilities, 
$\chi_s(\tau,\beta)=\langle T_{\tau}S_z(\tau)S_z(0) \rangle$ and 
$\chi_c(\tau,\beta)=\langle T_{\tau}\, \mbox{$:\!n(\tau)\!:$} \:
\mbox{$:\!n(0)\!:$}\rangle$, respectively, where
$S_z = \frac{1}{2}(n_{\uparrow} - n_{\downarrow})$, $\mbox{$:\!n\!:$} =
\sum_{\sigma}n_{\sigma}- \langle\sum_{\sigma}n_{\sigma}\rangle$,
and $\beta=1/T$ (taking $k_B\equiv 1$) plays the role of the system size.
The corresponding static susceptibilities follow from
$\chi_{c,s}(\beta) = \int_0^{\beta}\,d\tau\chi_{c,s}(\tau)$.  
Measuring powers of the local magnetization $\langle M_z^n \rangle =
\langle \big[\frac{1}{\beta}\int_0^{\beta}\,d\tau S_z(\tau) \big]^n\rangle$ 
allows construction of the Binder cumulant~\cite{Binder.81}
$B(U,\beta) = \langle M_z^4\rangle/\langle M_z^2\rangle^2$.
We supplement our $T>0$ (finite-$\beta$) CT-QMC results with static quantities
calculated arbitrarily close to $T=0$ ($\beta=\infty$) using the NRG method as
adapted to treat pseudogap impurity problems
\cite{Bulla.97,Buxton.98,Ingersent.02}.
NRG results presented below were obtained with Wilson discretization parameter
$\Lambda=9$, with $\Gamma_0$ corrected \cite{Buxton.98} to compensate for the
band discretization, and retaining all many-body states up to 50 times the
effective bandwidth of each iteration. 

\begin{figure}[t!]
\includegraphics[width=\linewidth]{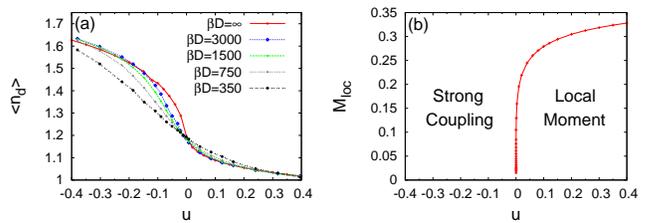}
\caption{Valence and local spin properties vs.\ $u=U/U_c -1$ for $r=0.6$,
$\Gamma_0=0.1D$, and $\varepsilon_d=-0.05D$:
(a) Occupancy $\langle n_d \rangle$ at the labeled temperatures \cite{note}.
The QCP ($u=0$) occurs at mixed valence, i.e., $\langle n_d \rangle \neq 1$.
(b) Local magnetization $M_{\mathrm{loc}}$, showing quenching of the
impurity spin for $u<0$ but the emergence of a free local moment for
$u > 0$.}
\label{Figure2}
\end{figure}

We focus our discussion on the representative case of a pseudogap exponent
$r=0.6$ with $\Gamma_0 = 0.1D$ and $\varepsilon_d = -0.05D$.
Figure \ref{Figure1} plots the variation of the Binder cumulant with $U$ at
different temperatures.
For small $U$, charge fluctuations are strong and the Binder cumulant lies
above the range $1<B(U,\beta)<3$ obeyed by a pure spin system~\cite{Glossop.11,Binder.81}; in this limit,
the low-energy behavior is close to that of a pseudogap resonant level.  
As $U$ increases, charge fluctuations are suppressed leading at low temperatures to
$B(U,\beta)<3$; this part of the strong-coupling phase exhibits a
true pseudogap Kondo effect.
For very large $U$, by contrast, $B(U,\beta)$ tends towards $1$ at low temperatures,
suggesting the presence of a decoupled impurity spin, characteristic of the 
local-moment phase.
We locate the phase boundary by the intersection of $B(U,\beta)$ curves for
different temperatures~\cite{Glossop.11} at $U_c/D = 0.06313 \pm 0.0008$. The NRG gives
$U_c = 0.06450D$, a small shift that can likely be attributed to residual
effects of NRG discretization.
The mixed-valent nature of the QCP is demonstrated in Fig.\ \ref{Figure2}(a),
where the local occupation $\langle n_d \rangle$ is seen to differ from unity
at $U=U_c$.
Note also that $\langle n_d \rangle$ displays significant temperature
dependence in the vicinity of the QCP.

\begin{figure}[t!]
\includegraphics[width=\linewidth]{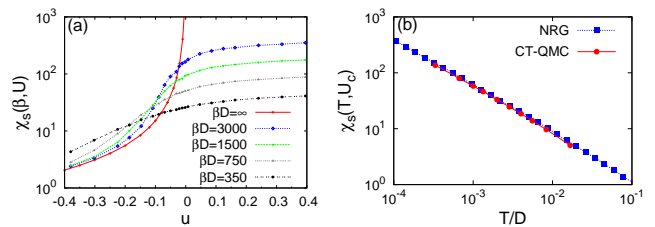}
\caption{Local static spin susceptibility $\chi_{s}(T,U)$
for $r=0.6$, $\Gamma_0=0.1D$, and $\varepsilon_d=-0.05D$,
(a) vs.\ $u=U/U_c\!-\!1$ at the labeled temperatures \cite{note}, and
(b) vs.\ $T$ at the critical point $U=U_c$.}
\label{Figure3}
\end{figure}

We are now in a position to look for a critical destruction of the Kondo effect
in this mixed-valent QCP, i.e., the continuous vanishing of an effective Kondo
energy scale signaled by the divergence of the zero-temperature static local
spin susceptibility $\chi_s$ as $U$ approaches $U_c$ from below. Such a
divergence is indeed seen in our zero-temperature $\chi_s$ vs.\ $U$ data
[Fig.\ \ref{Figure3}(a)] and in the temperature dependence of
$\chi_s$ at $U=U_c$ [Fig.\ \ref{Figure3}(b)].
Figure \ref{Figure2}(b) shows the $U$ dependence of the local magnetization
$M_{\mathrm{loc}} = \lim_{h\to 0} \lim_{T\to 0} \langle M_z \rangle$,
where $h$ is a local magnetic field
entering a term $h(n_{\uparrow}-n_{\downarrow})/2$ (with $g\mu_B\equiv 1$)
added to Eq.\ \eqref{ham}.
Since $M_{\mathrm{loc}}=0$ throughout the strong-coupling phase, and
$M_{\mathrm{loc}}$ rises continuously from zero on entry to the local-moment phase,
this quantity serves as an order parameter for the quantum phase transition.
Our results can be summarized as
\begin{eqnarray}
\label{spin_powers}
\chi_{s}(T,U=U_c) &\sim& T^{-x_{s}},
\nonumber \\
\chi_{s}(T=0,U < U_c) &\sim& |u|^{-\gamma_{s}},
\\
M_{\mathrm{loc}}(T=0,U>U_c) &\sim& u^{\beta_{s}},
\nonumber
\end{eqnarray}
where $u=U/U_c-1$. We find $x_s=0.80(3)$ from CT-QMC, in excellent agreement
with the NRG value $x_s=0.7908(3)$; the NRG also yields
$\gamma_{s}=1.42(2)$ and $\beta_s=0.1874(2)$.
These power-law behaviors are all defining
characteristics of critical Kondo destruction.

\begin{figure}[t!]
\includegraphics[width=\linewidth]{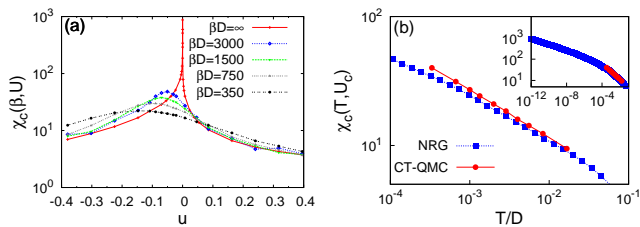}
\caption{Local static charge susceptibility $\chi_{c}(T,U)$ for $r=0.6$,
$\Gamma_0=0.1D$, and $\varepsilon_d=-0.05D$,
(a) vs.\ $u=U/U_c\!-\!1$ at the labeled temperatures \cite{note}, and
(b) vs.\ $T$ at the critical point $U=U_c$, where the discrepancy between
CT-QMC and NRG data is due mainly to the difference in $U_c$ values.
Inset: $\chi_{c}(T,U_c)$ over a wider range of $T$, showing the
slow crossover behavior.}
\label{Figure4}
\end{figure}

To probe valence fluctuations near the QCP, we turn to the static local charge
susceptibility $\chi_c(T,U)$. As shown in Fig.\ \ref{Figure4}(a),
\mbox{$\chi_c(T=0,U)$} increases with $U$ in the strong-coupling phase and
diverges as $U\to U_c^-$, in a manner similar to \mbox{$\chi_s(T=0,U)$}.
In the local-moment phase, the spin and charge responses are very different:
$\mbox{$\chi_s(T=0,U)$}=\infty$ but \mbox{$\chi_c(T=0,U)$} remains finite,
although it diverges as $U\to U_c^+$. In other words, the valence fluctuation
energy scale is nonzero in both phases, vanishing only when $U$ approaches
$U_c$ from either side. At $U=U_c$, $\chi_c$ has a singular temperature
dependence as shown in Fig.\ \ref{Figure4}(b). These behaviors
are consistent with
\begin{eqnarray}
\chi_{c}(T,U=U_c) &\sim& T^{-x_{c}},
\nonumber
\\
\chi_{c}(T=0,U) &\sim& |u|^{-\gamma_{c}}.
\end{eqnarray}
CT-QMC yields $x_c=0.36(3)$, while the NRG gives $x_c=0.120(1)$ (extracted at
temperatures much lower than can be accessed by CT-QMC) and
$\gamma_{c}=0.21(1)$. The difference between the two $x_c$ values stems from
a very slow crossover to the quantum critical regime [Fig.\
\ref{Figure4}(b) inset].
The much wider crossover window for $\chi_c$ compared with $\chi_s$
[Fig.\ \ref{Figure3}(b)] likely arises because $x_c<x_s$, meaning
that lower temperatures must be reached before sub-leading contributions to
$\chi_c$ become negligible.
We stress that the singularity of the charge susceptibility is unique to the 
mixed-valence QCP, and does not appear at its integer-valence counterpart.

We reach the important conclusions that the Kondo destruction occurs at a
genuinely mixed-valent QCP and that valence fluctuations are part of the
critical spectrum. Calculations for level energies $\varepsilon_d\ne
-0.05D$ (and hence different critical occupancies $\langle n_d\rangle$)
indicate that the critical exponents defined above depend on the band exponent
$r$, but not on the impurity valence. This implies that the divergence of the
static charge susceptibility is a universal property. At the same time, we find
that the critical behavior in the spin sector coincides 
with the model in its integer valence limit, i.e., the p-h-asymmetric pseudogap Kondo model
\cite{Ingersent.02}.  

We now discuss the dynamical scaling of $\chi_{s}(\tau,T)$ and 
$\chi_{c}(\tau,T)$. In analogy with the spin response at the
Kondo destruction QCP in the usual Kondo limit
\cite{Kirchner.08,Glossop.11}, we find that at $U=U_c$
both $\chi_{s}(\tau,T)$ and $\chi_{c}(\tau,T)$ 
collapse onto the conformal scaling form, showing
a power-law dependence on
$\pi T/\sin(\pi \tau T)$ with exponents $\eta_{c,s}$.
For the temperatures considered, the charge susceptibility has not yet reached
its asymptotic power-law behavior [based on Fig.\ \ref{Figure4}(b)]. Our results
thus imply that both leading and sub-leading terms of the critical
$\chi_c(\tau,T)$ scale in terms of $\pi T/\sin(\pi \tau T)$.
The scaling form means $\chi_s(\omega,T)$ and
$\chi_c(\omega,T)$ obey $\omega/T$ scaling~\cite{Glossop.11} at $U=U_c$.

\begin{figure}[t!]
\includegraphics[width=\linewidth]{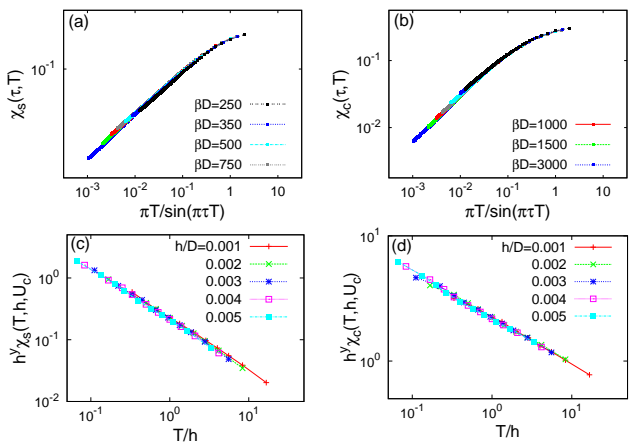}
\caption{Scaling of critical spin and charge responses for $r=0.6$,
$\Gamma_0=0.1D$, $\varepsilon_d=-0.05D$ and $U\simeq U_c$:
(a,b) Dynamical spin and charge susceptibilities vs.\ $\pi T/\sin(\pi \tau T)$. 
Both susceptibilities show excellent scaling collapse over two decades of
$\pi T/\sin(\pi \tau T)$, with $\eta_s = 0.20(3)$ and $\eta_c = 0.67(3)$.
(c,d) Static spin and charge susceptibilities vs.\ $T/h$ in the labeled
local magnetic fields $h$.}
\label{Figure5}
\end{figure}

We next consider the effect on the QCP of applying a finite local magnetic field
$h$. Consistent with the $\omega/T$ scaling we find $T/h$ scaling for fields
$|h|<T_K$, i.e.,
\begin{equation}
\chi_{c,s}(T,h,U_c) \sim h^{-y_{c,s}}f_{c,s}(T/h),
\label{scaleform2}
\end{equation}
where $f_{c,s}(x)$ is a scaling function [see Figs.\ \ref{Figure5}(c) and
\ref{Figure5}(d)].
Scaling collapse in $T/h$ further reflects the interacting nature of this
mixed-valence QCP.
At such an interacting QCP, critical exponents satisfy hyperscaling
relations that imply $y_{c,s}=x_{c,s}$, equalities
confirmed by our results to within numerical accuracy.

Our results are to be contrasted with the generalization of the
spin-density-wave QCP to the valence sector~\cite{Watanabe.10,Monthoux.04}. 
Like its spin counterpart~\cite{Hertz.76+Millis.93}, such a mixed-valent QCP
is Gaussian (noninteracting) and is not expected to obey either
energy-over-temperature or field-over-temperature scaling.

We now briefly consider the case $r=1$, motivated by critical Kondo screening
in graphene and d-wave superconductors. For $\Gamma_0=0.1D$ and
$\varepsilon_d=-0.05D$, we find $U_c/D = 0.05475 \pm 0.0006$ with CT-QMC, or
$0.05562$ with NRG. Both methods indicate that the local static charge
susceptibility diverges at $U_c$, along with the spin susceptibility. We find
$x_c$ to be an increasing function of $r$, but logarithmic corrections to
scaling \cite{Ingersent.02} prevent reliable determination of critical
exponents for $r=1$.

This work provides new insights into the unusual critical properties of
$\beta$-YbAlB$_4$ \cite{Okawa.10}, suggesting that Kondo destruction can
occur in this material even though it is mixed valent.
(Mixed valence is natural in this material given that its onset Kondo
temperature is high---on the order of 200\,K---and its mass enhancement is
moderate.)
The demonstration of $h/T$ scaling provides evidence that the experimentally
observed field-over-temperature scaling signals Kondo destruction. Our
finding of a rapid variation of $\langle n_d\rangle$ near the QCP, which can be
tested experimentally~\cite{note_nakatsuji}, suggests that the concentrated
lattice system is essentially quantum critical over a range of densities,
leading to the exciting possibility of quantum criticality occurring
over a region of parameter space rather than just at an isolated point.
Finally, our work raises intriguing questions about the extent to which
quantum-critical magnetic and valence degrees of freedom influence the
superconductivity observed in $\beta$-YbAlB$_4$.

From a general theoretical perspective, how quantum criticality can go beyond
the GLW framework of order-parameter fluctuations is a fundamental problem that
is important not only for heavy-fermion metals but also for QCPs arising in
insulating magnets and other strongly correlated systems. At present,
there are few concrete theoretical examples for such unconventional QCPs. By
identifying a new QCP in this category, our results provide another setting to
gain intuition about beyond-GLW QCPs in general.
 
 In summary, we have shown that mixed-valent quantum criticality can display the
phenomenon of Kondo destruction. The quantum critical point has a collapsing Kondo
energy scale and a singular charge-fluctuation spectrum.
The valence varies strongly with temperature near the critical point.
In the concentrated lattice case, similar quantum critical behavior is expected
to occur over an extended range of parameters. Our results raise the prospect of
unconventional quantum criticality in mixed-valent systems beyond
$\beta$-YbAlB$_4$.

We thank S.\ Nakatsuji, A.\ Nevidomskyy, S. Paschen, F. Steglich, H. Q. Yuan, 
and  L.\ Zhu for useful discussions.  
This work has been supported by NSF Grants No.\ DMR-0710540, DMR-1006985,
and DMR-1107814, and by Robert A.\ Welch Foundation Grant No.\ C-1411.
The calculations were in part performed on the Rice Computational Research
Cluster funded by the NSF and a partnership between Rice University, AMD and
Cray. We acknowledge the hospitality of the Max Planck Institutes for Chemical
Physics of Solids and Physics of Complex Systems (J.H.P., K.I.\ and Q.S.),
the Aspen Center for Physics (S.K.\ and Q.S.), 
and the Institute of Physics of Chinese Academy of Sciences (Q.S.)


\begin{thebibliography}{30}
\expandafter\ifx\csname natexlab\endcsname\relax\def\natexlab#1{#1}\fi
\expandafter\ifx\csname bibnamefont\endcsname\relax
  \def\bibnamefont#1{#1}\fi
\expandafter\ifx\csname bibfnamefont\endcsname\relax
  \def\bibfnamefont#1{#1}\fi
\expandafter\ifx\csname citenamefont\endcsname\relax
  \def\citenamefont#1{#1}\fi
\expandafter\ifx\csname url\endcsname\relax
  \def\url#1{\texttt{#1}}\fi
\expandafter\ifx\csname urlprefix\endcsname\relax\def\urlprefix{URL }\fi
\providecommand{\bibinfo}[2]{#2}
\providecommand{\eprint}[2][]{\url{#2}}

\bibitem{Mathur.98}
N.\ D.\ Mathur \textit{et al.}, Nature (London) \textbf{394}, 39 (1998).

\bibitem{Broun.08}
\bibinfo{author}{\bibfnamefont{D.\ M.} \bibnamefont{Broun}},
  \bibinfo{journal}{Nature Phys.} \textbf{\bibinfo{volume}{4}},
  \bibinfo{pages}{170} (\bibinfo{year}{2008}).

\bibitem{Sachdev}
S.\ Sachdev, {\it Quantum Phase Transitions}, CUP (1999), Cambridge.

\bibitem{Hertz.76+Millis.93}
\bibinfo{author}{\bibfnamefont{J. A.} \bibnamefont{Hertz}},
  \bibinfo{journal}{Phys. Rev. B} \textbf{\bibinfo{volume}{14}},
  \bibinfo{pages}{1165}, \bibinfo{year}{(1976)};
\bibinfo{author}{\bibfnamefont{A. J.} \bibnamefont{Millis}},
  \textit{ibid.} \textbf{\bibinfo{volume}{48}},
  \bibinfo{pages}{7183}, \bibinfo{year}{(1993)}.
  
\bibitem{Loehneysen.07+Si.10+Aronson.95+Schroeder.00+Paschen.04+Friedemann.10}
\bibinfo{author}{\bibfnamefont{H. v.} \bibnamefont{L\"{o}hneysen} \textit{et al.}},
 \bibinfo{journal}{Rev. Mod. Phys.} \textbf{\bibinfo{volume}{79}}, \bibinfo{pages}{1015}, (\bibinfo{year}{2007});
\bibinfo{author}{\bibfnamefont{Q.} \bibnamefont{Si}} \bibnamefont{and}
\bibinfo{author}{\bibfnamefont{F.} \bibnamefont{Steglich}} 
  \bibinfo{journal}{Science} \textbf{\bibinfo{volume}{329}},
  \bibinfo{pages}{1161} \bibinfo{year}{(2010)};
\bibinfo{author}{\bibfnamefont{M. C.} \bibnamefont{Aronson \textit{et al.}}},
  \bibinfo{journal}{Phys. Rev. Lett.} \textbf{\bibinfo{volume}{75}},
  \bibinfo{pages}{725} (\bibinfo{year}{1995});
\bibinfo{author}{\bibnamefont{A.} \bibnamefont{Schr\"oder \textit{et al.}}},
  \bibinfo{journal}{Nature} \textbf{\bibinfo{volume}{407}},
  \bibinfo{pages}{351} (\bibinfo{year}{2000});
\bibinfo{author}{\bibfnamefont{S.} \bibnamefont{Paschen \textit{et al.}}},
  \textit{ibid.} \textbf{\bibinfo{volume}{432}},
  \bibinfo{pages}{881} (\bibinfo{year}{2004});
\bibinfo{author}{\bibfnamefont{S.} \bibnamefont{Friedemann}},
 \bibinfo{journal}{Proc. Natl. Acad. Sci. USA}
 \textbf{\bibinfo{volume}{107}}, \bibinfo{pages}{14547} (\bibinfo{year}{2010}).

\bibitem{Park.06}
\bibinfo{author}{\bibfnamefont{T.} \bibnamefont{Park \textit{et al.}}}, 
  \bibinfo{journal}{Nature} \textbf{\bibinfo{volume}{440}},
  \bibinfo{pages}{65} (\bibinfo{year}{2006}).


\bibitem{Si.01+Si.03}
\bibinfo{author}{\bibfnamefont{Q.} \bibnamefont{Si \textit{et al.}}},
  \bibinfo{journal}{Nature} \textbf{\bibinfo{volume}{413}},
  \bibinfo{pages}{804} (\bibinfo{year}{2001});
  \bibinfo{journal}{Phys. Rev. B} \textbf{\bibinfo{volume}{68}},
  \bibinfo{pages}{115103} (\bibinfo{year}{2003}).


\bibitem[{\citenamefont{Coleman et al.}(2001)\citenamefont{Coleman, P\'{e}pin,
  Si, and Ramazashvili}}]{Coleman.01}
\bibinfo{author}{\bibfnamefont{P.} \bibnamefont{Coleman \textit{et al.}}},
  \bibinfo{journal}{J. Phys. Cond. Matt.} \textbf{\bibinfo{volume}{13}},
  \bibinfo{pages}{R723} (\bibinfo{year}{2001}).
  
  \bibitem{Shishido.05}
\bibinfo{author}{\bibfnamefont{H.} \bibnamefont{Shishido \textit{et al.}}}, 
  \bibinfo{journal}{J.~Phys.~Soc.~Jpn} \textbf{\bibinfo{volume}{74}},
  \bibinfo{pages}{1103} (\bibinfo{year}{2005}).

\bibitem{Yuan.03+Holmes.07+Rueff.11}
  H. Q. Yuan \textit{et al.}, Science \textbf{302}, 2104 (2003);
  A. T. Holmes, D. Jaccard, and K. Miyake,
    J.\ Phys.\ Soc.\ Jpn.\ \textbf{76}, 051002 (2007);
  J.-P. Rueff \textit{et al.},
    Phys.\ Rev.\ Lett.\ \textbf{106}, 186405 (2011).

\bibitem{Okawa.10}
\bibinfo{author}{\bibfnamefont{M.} \bibnamefont{Okawa \textit{et al.}}},
  \bibinfo{journal}{Phys. Rev. Lett.} \textbf{\bibinfo{volume}{104}},
  \bibinfo{pages}{247201} (\bibinfo{year}{2010}).
  
\bibitem{Nakatsuji.08}
\bibinfo{author}{\bibfnamefont{S.} \bibnamefont{Nakatsuji \textit{et al.}}},
  \bibinfo{journal}{Nature Phys.} \textbf{\bibinfo{volume}{4}},
  \bibinfo{pages}{603} (\bibinfo{year}{2008}).

\bibitem{Matsumoto.11}
\bibinfo{author}{\bibfnamefont{Y.} \bibnamefont{Matsumoto \textit{et al.}}},
  \bibinfo{journal}{Science} \textbf{\bibinfo{volume}{331}},
  \bibinfo{pages}{316} (\bibinfo{year}{2011}).


\bibitem{SmithSi-edmft}
J.\ L.\ Smith and Q.\ Si,
Phys. Rev. B{\bf 61}, 5184 (2000);
Q.\ Si and J.\ L.\ Smith,  
Phys. Rev. Lett. {\bf 77}, 3391 (1996).

\bibitem{Chitra} 
R.\ Chitra and G.\ Kotliar, 
Phys. Rev. Lett. {\bf 84}, 3678 (2000).

\bibitem[{\citenamefont{Ingersent and Si}(2002)}]{Ingersent.02}
\bibinfo{author}{\bibfnamefont{K.} \bibnamefont{Ingersent}} \bibnamefont{and}
  \bibinfo{author}{\bibfnamefont{Q.} \bibnamefont{Si}},
  \bibinfo{journal}{Phys. Rev. Lett.} \textbf{\bibinfo{volume}{89}},
  \bibinfo{pages}{076403} (\bibinfo{year}{2002}).

\bibitem[{\citenamefont{E. Gull et al}(2011)}]{Gull.11}
\bibinfo{author}{\bibfnamefont{E.} \bibnamefont{Gull \textit{et al.}}},
  \bibinfo{journal}{Rev. Mod. Phys.} \textbf{\bibinfo{volume}{83}},
  \bibinfo{pages}{349} (\bibinfo{year}{2011}).


\bibitem{Bulla.97}
  R. Bulla, Th. Pruschke, and A. C. Hewson,
  J. Phys.: Condens. Matter \textbf{9}, 10463 (1997).

\bibitem[{\citenamefont{Gonzalez-Buxton and Ingersent}(1998)}]{Buxton.98}
\bibinfo{author}{\bibfnamefont{C.} \bibnamefont{Gonzalez-Buxton}}
  \bibnamefont{and}
  \bibinfo{author}{\bibfnamefont{K.} \bibnamefont{Ingersent}},
  \bibinfo{journal}{Phys. Rev. B} \textbf{\bibinfo{volume}{57}},
  \bibinfo{pages}{14254} (\bibinfo{year}{1998}).
  
\bibitem{Chen.11+Jacob.11}
\bibinfo{author}{\bibfnamefont{J.-H.} \bibnamefont{Chen \textit{et al.}}},
 \bibinfo{journal}{Nature Phys.}
\textbf{\bibinfo{volume}{7}},\bibinfo{pages}{535} (\bibinfo{year}{2011});
\bibinfo{author}{\bibfnamefont{D.} \bibnamefont{Jacob}} \bibnamefont{and}
\bibinfo{author}{\bibfnamefont{G.} \bibnamefont{Kotliar}}
  \bibinfo{journal}{Phys. Rev. B} \textbf{\bibinfo{volume}{82}},
  \bibinfo{pages}{085423} (\bibinfo{year}{2010}).

\bibitem{Withoff.90+Vojta.02+Glossop.03+Fritz.06}
\bibinfo{author}{\bibfnamefont{D.} \bibnamefont{Withoff}} \bibnamefont{and}
  \bibinfo{author}{\bibfnamefont{E.} \bibnamefont{Fradkin}},
  \bibinfo{journal}{Phys. Rev. Lett.} \textbf{\bibinfo{volume}{64}},
  \bibinfo{pages}{1835} (\bibinfo{year}{1990});
\bibinfo{author}{\bibfnamefont{M.} \bibnamefont{Vojta}} \bibnamefont{and}
  \bibinfo{author}{\bibfnamefont{R.} \bibnamefont{Bulla}},
  \bibinfo{journal}{Phys. Rev. B} \textbf{\bibinfo{volume}{65}},
  \bibinfo{pages}{014511} (\bibinfo{year}{2001});
\bibinfo{author}{\bibfnamefont{M. T.} \bibnamefont{Glossop}} \bibnamefont{and}
  \bibinfo{author}{\bibfnamefont{D. E.} \bibnamefont{Logan}},
  \bibinfo{journal}{Europhys. Lett.} \textbf{\bibinfo{volume}{61}},
  \bibinfo{pages}{810} (\bibinfo{year}{2003}{\natexlab{a}});
\bibinfo{author}{\bibfnamefont{L.} \bibnamefont{Fritz}},
\bibinfo{author}{\bibfnamefont{S.} \bibnamefont{Florens}}, and
\bibinfo{author}{\bibfnamefont{M.} \bibnamefont{Vojta}},
  \bibinfo{journal}{Phys. Rev. B} \textbf{\bibinfo{volume}{74}},
  \bibinfo{pages}{144410} (\bibinfo{year}{2006}).

\bibitem[{\citenamefont{Vojta and Fritz}(2004)}]{Vojta.04}
\bibinfo{author}{\bibfnamefont{M.} \bibnamefont{Vojta}} \bibnamefont{and} 
  \bibinfo{author}{\bibfnamefont{L.} \bibnamefont{Fritz}},
  \bibinfo{journal}{Phys. Rev. B} \textbf{\bibinfo{volume}{70}},
  \bibinfo{pages}{094502} (\bibinfo{year}{2004}).

\bibitem{Glossop.11}
\bibinfo{author}{\bibfnamefont{M. T.} \bibnamefont{Glossop \textit{et al.}}},
  \bibinfo{journal}{Phys Rev. Lett.} \textbf{\bibinfo{volume}{107}},
  \bibinfo{pages}{076404} (\bibinfo{year}{2011}).

\bibitem[{\citenamefont{Binder}(1981)}]{Binder.81}
\bibinfo{author}{\bibfnamefont{K.} \bibnamefont{Binder}},
  \bibinfo{journal}{Z. Phys.~ } \textbf{\bibinfo{volume}{43}},
  \bibinfo{pages}{119} (\bibinfo{year}{1981}).
  
\bibitem{note}
  Data for $\beta=\infty$ were obtained using the NRG method. The
  remaining data represent CT-QMC results.

\bibitem[{\citenamefont{Kirchner et al.}(2008)\citenamefont{Kirchner and Si}}]{Kirchner.08}
\bibinfo{author}{\bibfnamefont{S.} \bibnamefont{Kirchner}} \bibnamefont{and}
\bibinfo{author}{\bibfnamefont{Q.} \bibnamefont{Si}},
  \bibinfo{journal}{Phys. Rev. Lett.} \textbf{\bibinfo{volume}{100}},
  \bibinfo{pages}{026403} (\bibinfo{year}{2008}).
  
\bibitem{Watanabe.10}
S.\ Watanabe and K.\ Miyake,
  J.\ Phys.\ Soc.\ Jpn.\ \textbf{79}, 033707 (2010).
  
\bibitem{Monthoux.04}
P.\ Monthoux and G.\ G.\ Lonzarich,
Phys.\ Rev.\ B \textbf{69}, 064517 (2004).

\bibitem{note_nakatsuji}
A strong temperature dependence in the quantum critical regime, similar
to that shown in Fig.~\ref{Figure2}(a), has been observed in recent
experimental measurements of the $4f$-valence of the Yb ions in
$\beta$-YbAlB$_4$ at low temperatures [S.~Nakatsuji, private
communication (2012)].

\end{thebibliography}
\end{document}